\documentclass[journal,onecolumn,12pt]{IEEEtran}
\usepackage{amsmath}
\usepackage{amsfonts}
\usepackage{amssymb}
\usepackage{graphicx}
\newtheorem{theorem}{Theorem}
\newtheorem{lemma}{Lemma}
\newtheorem{corollary}{Corollary}
\begin{document}

\title{Strong converse for the classical capacity of optical quantum communication channels}
\author{Bhaskar Roy Bardhan,
        Ra\'{u}l Garc\'{\i}a-Patr\'{o}n,
        Mark M.\ Wilde,
        and Andreas Winter
\thanks{B.~R.~Bardhan is with the Hearne Institute for Theoretical Physics and Department of Physics and Astronomy, Louisiana State University, Baton Rouge, Louisiana 70803, USA (e-mail: broyba1@tigers.lsu.edu).}
\thanks{R.~G-Patr\'{o}n is with the Center for Quantum Information and Communication, Ecole Polytechnique de Bruxelles, CP 165, Universite Libre de Bruxelles, 1050, Bruxelles, Belgium (email: raulgarciapatron@gmail.com).}
\thanks{M.~M.\ Wilde is with the Hearne Institute for Theoretical Physics, Department of Physics and Astronomy, Louisiana State University, Baton Rouge, Louisiana 70803, USA, and
Center for Computation and Technology, Louisiana State University, Baton Rouge, Louisiana 70803, USA (email: mwilde@gmail.com).}
\thanks{A.~Winter is with ICREA \& F\'isica Te\'orica, Informaci\'o i Fenomens Qu\'antics, Universitat Aut\`{o}noma de Barcelona, ES-08193, Bellaterra (Barcelona), Spain (email: andreas.winter@uab.cat).}
\thanks{B.~R.~Bardhan and M.~M.\ Wilde are grateful for support from the Department of Physics and Astronomy, Louisiana State University. M.~M.\ Wilde was also supported by the DARPA Quiness Program through US Army Research
Office award W31P4Q-12-1-0019. A.~Winter acknowledges financial support by the Spanish MINECO, project FIS2008-01236
with the support of FEDER funds, the EC STREP \textquotedblleft
RAQUEL\textquotedblright, the ERC Advanced Grant \textquotedblleft
IRQUAT\textquotedblright, and the Philip Leverhulme Trust. R.~G-Patr\'{o}n acknowledges support from
 a Postdoctoral Researcher fellowship of the F.R.S.-FNRS and the action Back to Belgium Grants 
of the Belgian Federal Science Policy. R.~G-Patr\'{o}n also acknowledges
financial support from the F.R.S.-FNRS under
projects T.0199.13 and HIPERCOM, as well as from the Interuniversity Attraction Poles program 
of the Belgian Science Policy Office under Grant No.~IAP P7-35 Photonics@be.}
\thanks{This paper was presented in part at the IEEE International Symposium on Information Theory (ISIT) (2014).}
\thanks{Copyright \textcircled{c} 2014 IEEE. Personal use of this material is permitted.}}

\maketitle

\begin{abstract}
We establish the classical capacity of optical quantum channels as a sharp transition between 
two regimes---one which is an error-free regime for communication rates below the capacity, and 
the other in which the probability of correctly decoding a classical message converges exponentially
fast to zero if the communication rate exceeds the classical capacity. This result is obtained by proving a 
strong converse theorem for the classical capacity of all phase-insensitive bosonic Gaussian channels, a well-established model of optical
quantum communication channels, such as lossy optical fibers, amplifier and free-space communication. The theorem holds
under a particular photon-number occupation constraint, which
we describe in detail in the paper.
Our result bolsters the understanding of the classical 
capacity of these channels and opens the path to applications, such as proving the security of
 noisy quantum storage models of cryptography with optical links.
\end{abstract}


\section{Introduction}

One of the most fundamental tasks in quantum information theory is to
determine the ultimate limits on achievable data transmission rates for a
noisy communication channel. The classical capacity is defined as the maximum
rate at which it is possible to send classical data over a quantum channel
such that the error probability decreases to zero in the limit of many
independent uses of the channel \cite{Hol98,SW97}. As such, the classical
capacity serves as a distinctive bound on our ability to faithfully recover
classical information sent over the channel.

The above definition of the classical capacity states that (a) for any rate
below capacity, one can communicate with vanishing error probability in the limit of many channel
uses and (b) there cannot exist such a communication scheme in the
limit of many channel uses whenever the rate exceeds the capacity. However,
strictly speaking, for any rate $R$ above capacity, the above definition
leaves open the possibility for one to increase the communication rate $R$ by
allowing for some error $\varepsilon>0$. Leaving room for the possibility of
such a trade-off between the rate $R$ and the error $\varepsilon$ is the
characteristic of a ``weak converse,'' and the corresponding capacity is
sometimes called the weak capacity. A \emph{strong converse}, on the contrary,
establishes the capacity as a very sharp threshold, so that there is no such
room for a trade-off between rate and error in the limit of many independent
uses of the channel. That is, it guarantees that the error probability of any
communication scheme asymptotically converges to one if its rate exceeds the
classical capacity.

Despite their significance in understanding the ultimate information-carrying
capacity of noisy communication channels, strong converse theorems are known
to hold only for a handful of quantum channels: for those with classical
inputs and quantum outputs \cite{Ogawa,W99} (see earlier results for all
classical channels \cite{Wolfowitz1964,Arimoto}), for all covariant channels
with additive minimum output R\'{e}nyi entropy \cite{KW09}, for all
entanglement-breaking and Hadamard channels \cite{Entanglementbreaking}, as
well as for pure-loss bosonic channels \cite{StrongConversePureLoss}.

In this paper, we consider the encoding of classical messages into optical quantum states 
and the  transmission of these codewords over phase-insensitive Gaussian channels.
Phase-insensitive Gaussian channels are invariant with respect to phase-space
rotations \cite{PBLSC12,CD94,HW01,ISS11}, and they are considered to be one of the most
practically relevant models to describe free space or optical fiber
transmission, or transmission of classical messages through dielectric media,
etc. In fact, phase-insensitive Gaussian channels constitute a broad class of
noisy bosonic channels, encompassing all of the following: thermal noise
channels (in which the signal photon states are mixed with a thermal state),
additive noise channels (in which the input states are randomly displaced in
phase space), and noisy amplifier channels
\cite{PBLSC12,GHG13,GPCH13,PracticalPurposes}. We prove that a
strong converse theorem holds for the classical capacity of these channels,
when imposing a photon-number occupation constraint on the inputs of the channel.

In some very recent studies \cite{GHG13,MGH13,GPCH13}, a solution to the
long-standing minimum output entropy conjecture \cite{HW01,GGLMS04} has been
established for all phase-insensitive Gaussian channels, demonstrating that
the minimum output entropy for such channels is indeed achieved by the vacuum
input state. The major implication of this work is that we now know the
classical capacity of any phase-insensitive Gaussian channel whenever there is
a mean photon-number constraint on the channel inputs (the capacity otherwise
being infinite if there is no photon number constraint). For instance,
consider the thermal noise channel represented by a beamsplitter with
transmissivity $\eta\in\lbrack0,1]$ mixing signaling photons (with mean
photon number $N_{S}$) with a thermal state of mean photon number $N_{B}$.
The results in \cite{GHG13,MGH13,GPCH13} imply that the classical capacity of this channel
is given by
\begin{equation}
g(\eta N_{S}+(1-\eta)N_{B})-g((1-\eta)N_{B}),
\end{equation}
where $g(x)\equiv(x+1)\log_{2}(x+1)-x\log_{2}x$ is the entropy of a bosonic
thermal state with mean photon number $x$. However, the corresponding converse
theorem, which can be inferred as a further implication of their work, is only
a weak converse, in the sense that the upper bound on the communication rate
$R$ of any coding scheme for the thermal noise channel can be written in the
following form:
\begin{equation}
R\leq\frac{1}{1-\varepsilon}[g(\eta N_{S}+(1-\eta)N_{B})-g((1-\eta
)N_{B})+h_{2}(\varepsilon)],
\end{equation}
where $\varepsilon$ is the error probability, and $h_{2}(\varepsilon)$ is the
binary entropy with the property that $\text{lim}_{\varepsilon\rightarrow
0}h_{2}(\varepsilon)=0$. That is, only in the limit $\varepsilon\rightarrow0$
does the above expression serve as the classical capacity of the channel,
leaving room for a possible trade-off between rate and error probability. This
observation also applies to the classical capacity of all other
phase-insensitive Gaussian channels mentioned above.

In the present work, we prove that a strong converse theorem holds for the
classical capacity of all phase-insensitive Gaussian channels when imposing
a photon-number occupation constraint. This means that if we demand that the
average code density operator for the codewords, which are used for
transmission of classical messages, is constrained to have a large shadow onto
a subspace with photon number no larger than some fixed amount, then the
probability of successfully decoding the message converges to zero in the
limit of many channel uses if the rate $R$ of communication exceeds the
classical capacity of these channels. We provide a mathematical definition
in \eqref{Constraint}.

This paper is structured as follows. In Section~\ref{Prelims}, we review
several preliminary ideas, including some definitions and notation for
phase-insensitive Gaussian channels, and we recall structural decompositions
of them that we exploit in our proof of the strong converse. We also recall
the definition of the quantum R\'enyi entropy and an inequality that relates
it to the smooth min-entropy \cite{RWISIT1}. In Section~\ref{SCT}, we then
prove our main result, i.e., that the strong converse holds for the classical
capacity of all phase-insensitive Gaussian channels when imposing a
photon-number occupation constraint. We conclude with a brief summary in
Section~\ref{sec:conclusion} along with suggestions for future research.

\section{Preliminaries}

\label{Prelims}

\subsection{Phase-insensitive Gaussian channels}

\label{sec:bosonic-Gaussian} Bosonic Gaussian channels play a key role in
modeling optical communication channels, such as optical fibers or free space
transmission. They are represented by completely positive and trace preserving
(CPTP) maps evolving Gaussian input states into Gaussian output
states~\cite{WPGCRSL12,symplectic,CD94}. (A Gaussian state is completely
characterized by a mean vector and a covariance matrix \cite{WPGCRSL12}.)
Single-mode Gaussian channels are characterized by two matrices $X$ and $Y$
acting on the covariance matrix $\Gamma$ of a single-mode Gaussian state in
the following way:%
\begin{equation}
\Gamma\longrightarrow\Gamma^{\prime}=X\Gamma X^{T}+Y,
\end{equation}
where $X^{T}$ is the transpose of the matrix $X$. Here $X$ and $Y$ are both
$2\times2$ real matrices, satisfying
\begin{equation}
Y\geq0,\ \ \ \ \ \ \ \det Y\geq(\det X-1)^{2},
\end{equation}
in order for the map to be a legitimate completely positive trace preserving
map (see \cite{WPGCRSL12} for more details). A bosonic Gaussian quantum
channel is said to be \textquotedblleft quantum-limited' if the inequality
above (involving $\det X$ and $\det Y$) is saturated~\cite{Caves82,ISS11,GHG13,GPCH13}. 
For instance, phase-insensitive Gaussian channels are quantum-limited when
 their environment is initially in a vacuum state.

In this work, we are interested in the most physically relevant set of 
phase-insensitive channels that corresponds to the choice
\begin{align}
X &  =\text{diag}\left(  \sqrt{\tau},\sqrt{\tau}\right)
,\label{eq:phase-insensitive-Gaussian}\\
Y &  =\text{diag}\left(  \nu,\nu\right)  ,\nonumber
\end{align}
with $\tau,\nu\geq0$ obeying the constraint above. The action of such
phase-insensitive channels on an input signal mode can be uniquely described
by their transformation of the symmetrically ordered characteristic function,
defined as
\begin{equation}
\chi(\mu)\equiv\operatorname{Tr}[\rho D(\mu)],
\end{equation}
where $D(\mu)\equiv\exp(\mu\hat{a}^{\dagger}-\mu^{\ast}\hat{a})$ is the
displacement operator for the input signal mode $\hat{a}$ \cite{WPGCRSL12}.
For the Gaussian channels, the transformed characteristic function at the
output is given by \cite{GPCH13,GHG13,symplectic}
\begin{equation}\chi^{\prime}(\mu)=\chi(\sqrt{\tau}\mu)\exp(-\nu\left\vert
\mu\right\vert ^{2}/2) .
\end{equation} 

\subsubsection{Examples}

The canonical phase-insensitive Gaussian channels are the thermal noise
channel, the additive noise channel, and the amplifier channel
\cite{PracticalPurposes,WPGCRSL12,GHG13,MGH13,PBLSC12,SKPPC13,GGLMS04,GPCH13}.

The thermal channel $\mathcal{E}_{\eta,N_{B}}$ can be represented by a
beamsplitter of transmissivity $\eta\in[0,1]$ that couples the input signal of
mean photon number $N_{S}$ with a thermal state of mean photon number $N_{B}$.
The special case $N_{B}=0$ corresponds to the pure-loss bosonic channel
$\mathcal{E}_{\eta}$, where the state injected by the environment is the
vacuum state.

In the additive noise channel $\mathcal{N}_{\bar{n}}$, each signal mode is
randomly displaced in phase space according to a Gaussian distribution. The
additive noise channel $\mathcal{N}_{\bar{n}} $ is completely characterized by
the variance $\bar{n}$ of the noise introduced by the channel.

The quantum amplifier channel $\mathcal{A}_{G}^{N}$ is characterized by its
gain $G \geq1$ and the mean number of photons $N$ in the associated environment
input mode (which is in a thermal state). The amplifier channel 
$\mathcal{A}_{G}^{N}$ is quantum-limited when the environment is in 
the vacuum state (we will denote such a quantum-limited
amplifier by $\mathcal{A}_{G}^{0}$).

The transformed characteristic functions for these Gaussian channels are given
by \cite{MGH13,GGLMS04,PracticalPurposes}%

\begin{equation}
\chi^{\prime}(\mu)=
\begin{cases}
\chi(\sqrt{\eta}\mu) e^{-(1-\eta)(N_{B}+1/2)|\mu|^{2}}~ & \text{for}%
~\mathcal{E}_{\eta,N_{B}}\\
\chi(\mu) e^{-\bar{n} |\mu|^{2}}~ & \text{for}~\mathcal{N}_{\bar{n}}\\
\chi(\sqrt{G} \mu) e^{-(G-1)(N+1/2)|\mu|^{2}}~ & \text{for}~\mathcal{A}%
_{G}^{N}.
\end{cases}
\end{equation}

\subsubsection{Structural decompositions}

Using the composition rule of Gaussian bosonic channels \cite{CGH06}, any
phase-insensitive Gaussian bosonic channel (let us denote it by $\mathcal{P}$)
can be written as a concatenation of a pure-loss channel followed by a
quantum-limited amplifier \cite{PBLSC12}
\begin{equation}
\mathcal{P}=\mathcal{A}_{G}^{0}\circ\mathcal{E}_{T},\label{decomposition}%
\end{equation}
where $\mathcal{E}_{T}$ is a pure-loss channel with parameter $T\in\left[
0,1\right]  $ and $\mathcal{A}_{G}^{0}$ is a quantum-limited amplifier with
gain $G\geq1$, these parameters chosen so that $\tau=TG$ and $\nu=G\left(
1-T\right)  +G-1$ (with $\tau$ and $\nu$ defined in
(\ref{eq:phase-insensitive-Gaussian})).

For instance, the additive noise channel $\mathcal{N}_{\bar{n}}$ can be
realized as a pure-loss channel with transmissivity $T=1/(\bar{n}+1)$ followed
by a quantum-limited amplifier channel with gain $G=\bar{n}+1$. Also, we can
consider the thermal noise channel $\mathcal{E}_{\eta,N_{B}}$ as a cascade of
a pure-loss channel with transmissivity ${T}=\eta/G$ followed by a
quantum-limited amplifier channel with gain $G=(1-\eta)N_{B}+1$. These two
observations are equivalent to
\begin{align}
\mathcal{N}_{\bar{n}}(\rho)  &  =(\mathcal{A}_{\bar{n}+1}^{0}\circ
\mathcal{E}_{\frac{1}{\bar{n}+1}})(\rho),\label{additive-decomposition}\\
\mathcal{E}_{\eta,N_{B}}(\rho)  &  =(\mathcal{A}_{(1-\eta)N_{B}+1}^{0}%
\circ\mathcal{E}_{T})(\rho). \label{thermal-decomposition}%
\end{align}
The above structural decompositions are useful in establishing the classical
capacity as well as the minimum output entropy for all phase-insensitive
channels \cite{GHG13,MGH13,GPCH13}.

\subsubsection{Classical capacitites of phase-insensitive channels}

Holevo, Schumacher, and Westmoreland (HSW) characterized the classical
capacity of a quantum channel $\mathcal{N}$ in terms of a quantity now known
as the Holevo information~\cite{Hol98,SW97}
\begin{equation}
\chi(\mathcal{N})\equiv\max_{\{p_{X}(x),\rho_{x}\}}I(X;B)_{\rho},
\end{equation}
where $\{p_{X}(x),\rho_{x}\}$ represents an ensemble of quantum states, and
the quantum mutual information $I(X;B)_{\rho}\equiv H(X)_{\rho}+H(B)_{\rho
}-H(XB)_{\rho}$, is defined with respect to a classical-quantum state
$\rho_{XB}\equiv\sum_{x}p_{X}\left(  x\right)  \left\vert x\right\rangle
\left\langle x\right\vert _{X}\otimes\mathcal{N}\left(  \rho_{x}\right)  _{B}%
$. The above formula given by HSW for certain quantum channels is  additive whenever
\begin{equation}
\chi(\mathcal{N}^{\otimes n})=n\chi(\mathcal{N}),
\end{equation}
for any positive integer $n$. For such quantum channels, the HSW formula is indeed
equal to the classical capacity of those channels. However, a regularization is
thought to be required in order to characterize the classical capacity of quantum channels
for which the HSW formula cannot be shown to be additive. The classical
capacity in general is then characterized by the following regularized
formula:%
\begin{equation}
\chi_{\text{reg}}(\mathcal{N})\equiv\lim_{n\rightarrow\infty}\frac{1}%
{n}\chi(\mathcal{N}^{\otimes n}).
\end{equation}

The recent breakthrough works in \cite{GHG13,GPCH13} (along with earlier results in
\cite{HW01,GGLMSY04}) have established the following expressions for the classical
capacities of various phase-insensitive channels:
\begin{align}
C(\mathcal{E}_{\eta,N_{B}}) &  =g(\eta N_{S}+(1-\eta)N_{B})-g((1-\eta
)N_{B}),\label{thermalcapacity}\\
C(\mathcal{N}_{\bar{n}}) &  =g(N_{S}+\bar{n})-g(\bar{n}%
)\,,\label{additivecapacity}\\
C(\mathcal{A}_{G}^{N}) &  =g(GN_{S}%
+(G-1)(N+1))-g((G-1)(N+1)),\label{amplifiercapacity}%
\end{align}
where $N_{S}$ is the mean input photon number. In general, the classical
capacity of any phase-insensitive Gaussian channel is equal to%
\begin{equation}
g (N_{S}^{\prime})-g (N_{B}^{\prime}),
\label{diff-equation}
\end{equation}
where $N_{S}^{\prime}=\tau N_{S}+\left(  \tau+\nu-1\right)  /2$ and $N_{B}^{\prime}= \left(
\tau+\nu-1\right)  /2$, with $\tau$ and $\nu$
defined in (\ref{eq:phase-insensitive-Gaussian}).
In the above, $N_{S}^{\prime}$ is equal to the mean number of photons at the output 
when a thermal state of mean photon number $N_S$ is input, and
$N_{B}^{\prime}$ is equal to the mean number of noise photons when the vacuum state is sent in.
Note that the capacities in (\ref{thermalcapacity}), (\ref{additivecapacity}),
and (\ref{amplifiercapacity}) all have this particular form
(but they differ in the corresponding mean number of photons).
The classical capacities specified above are attainable by
using coherent-state encoding schemes for the respective channels \cite{HW01}.
We will show in Section~\ref{SCT} that these expressions can also be
interpreted as strong converse rates.

\subsection{Quantum R\'enyi entropy and smooth min-entropy}

The quantum R\'{e}nyi entropy $H_{\alpha}(\rho)$ of a density operator $\rho$
is defined for $0<\alpha<\infty$, $\alpha\neq1$ as
\begin{equation}
H_{\alpha}(\rho)\equiv\frac{1}{1-\alpha}\log_{2}\operatorname{Tr}[\rho
^{\alpha}]\,.
\end{equation}
It is a monotonic function of the \textquotedblleft$\alpha$%
-purity\textquotedblright\ $\operatorname{Tr}[\rho^{\alpha}]$, and the von
Neumann entropy $H(\rho)$ is recovered from it in the limit $\alpha
\rightarrow1$:%
\begin{equation}
\lim_{\alpha\rightarrow1}H_{\alpha}(\rho)=H(\rho)\equiv-\operatorname{Tr}%
[\rho\log_{2}\rho]\,.
\end{equation}
The min-entropy is recovered from it in the limit as $\alpha\rightarrow\infty
$:%
\begin{equation}
\lim_{\alpha\rightarrow\infty}H_{\alpha}(\rho)=H_{\min}\left(  \rho\right)
\equiv-\log_{2}\left\Vert \rho\right\Vert _{\infty},
\end{equation}
where $\left\Vert \rho\right\Vert _{\infty}$ is the infinity norm of $\rho$.

The quantum R\'enyi entropy of order $\alpha> 1$ of a thermal state with mean
photon number $N_{B}$ can be written as~\cite{GLMS04}%
\begin{equation}
\frac{\log_{2}\left[  \left(  N_{B}+1\right)  ^{\alpha}-N_{B}^{\alpha}\right]
}{\alpha-1}.%
\end{equation}
For an additive noise channel $\mathcal{N}_{\bar{n}}$, the R\'{e}nyi entropy
$H_{\alpha}(\mathcal{N}_{\bar{n}}(\rho))$ for $\alpha> 1$ achieves its minimum
value when the input $\rho$ is the vacuum state $\vert0\rangle\langle0\vert
$~\cite{MGH13}:
\begin{equation}
\min_{\rho}H_{\alpha}(\mathcal{N}_{\bar{n}}(\rho))=H_{\alpha}(\mathcal{N}%
_{\bar{n}}(|0\rangle\langle0|))=\frac{\log_{2}[(\bar{n}+1)^{\alpha}-\bar
{n}^{\alpha}]}{\alpha-1}\,~\text{for}~\alpha> 1.
\end{equation}
Similarly, for the thermal noise channel $\mathcal{E}_{\eta,N_{B}}$, the
R\'{e}nyi entropy $H_{\alpha}(\mathcal{E}_{\eta,N_{B}}(\rho))$ for $\alpha> 1$
achieves its minimum value when the input $\rho$ is the vacuum state
$\vert0\rangle\langle0 \vert$~\cite{MGH13}:
\begin{equation}
\min_{\rho}H_{\alpha}(\mathcal{E}_{\eta,N_{B}}(\rho))=H_{\alpha}%
(\mathcal{E}_{\eta,N_{B}}(|0\rangle\langle0|))=\frac{\log_{2}[((1-\eta
)N_{B}+1)^{\alpha}-((1-\eta)N_{B})^{\alpha}]}{\alpha-1}\,~\text{for}~\alpha>
1.
\end{equation}
In general, the main result of \cite{MGH13} shows that the minimum output
R\'enyi entropy of any phase-insensitive Gaussian channel $\mathcal{P}$ is
achieved by the vacuum state:
\begin{equation}
\min_{\rho^{(n)}}H_{\alpha}(\mathcal{P}^{\otimes n}(\rho^{(n)}))= n H_{\alpha
}(\mathcal{P}(\vert0 \rangle\langle0 \vert) . \label{eq:main-result-MGH}%
\end{equation}

The above definition of the R\'{e}nyi entropy can be generalized to the smooth
R\'{e}nyi entropy. This notion was first introduced by Renner and Wolf for
classical probability distributions \cite{RWISIT1} and was later generalized
to the quantum case (density operators). For a given density operator $\rho$,
one can consider the set $\mathcal{B}^{\varepsilon}(\rho)$ of density
operators $\tilde{\rho}$ that are $\varepsilon$-close to $\rho$ in trace
distance for $\varepsilon\geq0$~\cite{RennerThesis}. The $\varepsilon$-smooth
quantum R\'{e}nyi entropy of order $\alpha$ of a density operator $\rho$ is
defined as~\cite{RennerThesis}%
\begin{equation}
H_{\alpha}^{\varepsilon}(\rho)\equiv\left\{
\begin{array}
[c]{cc}%
\inf_{\tilde{\rho}\in\mathcal{B}^{\varepsilon}(\rho)}H_{\alpha}(\tilde{\rho
}) & 0\leq\alpha<1\\
\sup_{\tilde{\rho}\in\mathcal{B}^{\varepsilon}(\rho)}H_{\alpha}(\tilde{\rho
}) & 1<\alpha<\infty
\end{array}
\right.  .
\end{equation}
In the limit as $\alpha\rightarrow\infty$, we recover the smooth min-entropy
of $\rho$ \cite{RennerThesis,Tomamichelthesis}:%
\begin{equation}
H_{\min}^{\varepsilon}(\rho)\equiv\sup_{\tilde{\rho}\in\mathcal{B}%
^{\varepsilon}(\rho)}\left[  -\log_{2}\left\Vert \widetilde{\rho}\right\Vert
_{\infty}\right]  \,.\label{normsmoothing}%
\end{equation}
From the above, we see that the following relation holds%
\begin{equation}
\inf_{\widetilde{\rho}\in\mathcal{B}^{\varepsilon}\left(  \rho\right)
}\left\Vert \widetilde{\rho}\right\Vert _{\infty}=2^{-H_{\min}^{\varepsilon
}\left(  \rho\right)  }\,.
\end{equation}
A relation between the smooth min-entropy and the R\'{e}nyi entropy of order
$\alpha>1$ is given by the following inequality \cite{RWISIT1}%
\begin{equation}
H_{\min}^{\varepsilon}\left(  \rho\right)  \geq H_{\alpha}\left(  \rho\right)
-\frac{1}{\alpha-1}\log_{2}\left(  \frac{1}{\varepsilon}\right)
.\label{Renyismoothing}%
\end{equation}
We will use this relation, along with the minimum output entropy results from
\cite{MGH13}, to prove the strong converse theorem for the classical capacity
of all phase-insensitive Gaussian channels.

\section{Strong converse for all phase-insensitive Gaussian channels}

\label{SCT}In this section, we consider the transmission of classical messages
through phase-insensitive channels and show that a strong converse theorem
holds for the classical capacity of these channels. Before doing so, we first
make the following two observations:

\begin{itemize}
\item If the input signal states are allowed to have an arbitrarily large
number of photons, then the classical capacity of the corresponding channel is
infinite \cite{HW01}. Thus, in order to have a sensible notion of the
classical capacity for any quantum channel, one must impose some kind of
constraint on the photon number of the states being fed into the channel. The
most common kind of constraint is to demand that the mean number of photons in
any signal transmitted through the channel can be at most $N_{S}\in
\lbrack0,\infty)$. This is known as the mean photon number constraint and is
commonly used in establishing the information-carrying capacity of a given
channel \cite{HW01,GGLMSY04,GHG13,GPCH13}. 
However, following the same arguments as
in \cite{StrongConversePureLoss} (and later in \cite{BW13}), we can show that
the strong converse need not hold under such a mean photon number constraint.
Indeed, as detailed in \cite{StrongConversePureLoss},
there exists a communication scheme which allows for trading
between communication rate and success probability, 
excluding the possibility of a strong converse holding
under a mean photon number constraint.
So instead, we prove that the strong converse theorem holds under a 
photon-number occupation constraint (see below for the definition and 
implication of this constraint) on the number of photons in the input states.

\item Our proof of the strong converse theorem for the phase-insensitive
channels can be regarded as a generalization of the arguments used in
establishing the strong converse theorem for the classical capacity of the
noiseless qubit channel~\cite{N99,KW09}. However, a comparison of our proof
here and that for the noiseless qubit channel reveals that it is a significant
generalization. Furthermore, our proof here also invites comparison with the
proof of the strong converse for covariant channels with additive minimum
output R\'enyi entropy~\cite{KW09}, especially since additivity of minimum
output R\'enyi entropies plays a critical role in the present paper.
\end{itemize}

Let $\rho_{m}$ denote a codeword of any code for communication over the
phase-insensitive Gaussian channel~$\mathcal{P}$. The \emph{photon-number occupation constraint} 
that we impose on the codebook is to require that the
average code density operator $\frac{1}{M}\sum_{m}\rho_{m}$ ($M$ is the total
number of messages) has a large shadow onto a subspace with photon number no
larger than some fixed amount $nN_{S}$. Such a constraint on the channel inputs 
can be defined by introducing a photon number cutoff projector $\Pi_{L}$ that projects onto a
subspace of $n$ bosonic modes such that the total photon number is no larger
than $L$:
\begin{equation}
\Pi_{L}\equiv\sum_{a_{1},\ldots,a_{n}:\sum_{i}a_{i}\leq L}|a_{1}\rangle\langle
a_{1}|\otimes\ldots\otimes|a_{n}\rangle\langle a_{n}|,
\end{equation}
where $|a_{i}\rangle$ is a photon number state of photon number $a_{i}$.
The rank of the  above projector $\Pi_{\lceil nN_{S}\rceil}$ has been shown to be never larger than
$2^{n[g(N_{S})+\delta_{0}]}$ (Lemma 3 in \cite{StrongConversePureLoss}), i.e.,%
\begin{equation}
\text{Tr}\left\{  \Pi_{\lceil nN_{S}\rceil}\right\}  \leq2^{n[g(N_{S}%
)+\delta_{0}]}, \label{Rank}%
\end{equation}
where $\delta_{0} \geq\frac{1}{n}(\log_{2} e+\log_{2}(1+\frac{1}{N_{S}}))$, so that
$\delta_{0}$ can be chosen arbitrarily small by taking $n$ large enough.

Mathematically, the photon-number occupation constraint can then be written as
\begin{equation}
\frac{1}{M}\sum_{m}\text{Tr}\left\{  \Pi_{\left\lceil nN_{S}\right\rceil }%
\rho_{m}\right\}  \geq1-\delta_{1}(n), \label{Constraint}%
\end{equation}
where $\delta_{1}(n)$ is a function that decreases to zero as $n$ increases.
In fact, the coherent-state encodings that attain the known capacities of the
phase-insensitive channels do indeed satisfy the photon-number occupation constraint, with an exponentially decreasing $\delta_{1}(n)$, if coherent
states with mean photon number per mode $< N_{S}-\delta$ are used, with
$\delta$ being a small positive number (see Ref.~\cite{StrongConversePureLoss}
for an argument along these lines).

The first important step in proving the strong converse is to show that if
most of the probability mass of the input state of the phase-insensitive
channel $\mathcal{P}$ is in a subspace with photon number no larger than
$nN_{S}$, then most of the probability mass of the channel output is in a
subspace with photon number no larger than $nN_{S}^{\prime}$, where
$N_{S}^{\prime}$ is the mean energy of the output state. We state this as the
following lemma:

\begin{lemma}
\label{lem:photon-num-lemma}Let $\rho^{(n)}$ denote a density operator on $n$
modes that satisfies%
\begin{equation}
\operatorname{Tr}\{\Pi_{\lceil nN_{S}\rceil}\rho^{(n)}\} \geq1-\delta_{1}(n),
\end{equation}
where $\delta_{1}(n)$ is defined in \eqref{Constraint}. Let $\mathcal{P}$ be a
phase-insensitive Gaussian channel with parameters $\tau$ and $\nu$ as defined
in \eqref{eq:phase-insensitive-Gaussian}. Then%
\begin{equation}
\operatorname{Tr}\{\Pi_{\lceil nN_{S}^{\prime}+\delta_{2})\rceil}%
\mathcal{P}^{\otimes n}(\rho^{(n)})\}\geq1-\delta_{1}(n)-2\sqrt{\delta_{1}%
(n)}-\delta_{3}(n),
\end{equation}
where $N_{S}^{\prime}=\tau N_{S}+\left(  \tau+\nu-1\right)  /2$,
$\mathcal{P}^{\otimes n}$ represents $n$ instances of $\mathcal{P}$ that act
on the density operator $\rho^{(n)}$, $\delta_{2}$ is an arbitrarily small
positive constant, and $\delta_{3}(n)$ is a function of $n$ decreasing to zero
as $n\rightarrow\infty$.
\end{lemma}

\begin{IEEEproof}
The proof of this lemma is essentially the same as the proof of Lemma~1 of
\cite{BW13}, with some minor modifications. We include the details of it for
completeness. We first recall the structural decomposition in
(\ref{decomposition}) for any phase-insensitive channel:
\begin{equation}
\mathcal{P}(\rho)=\left(  \mathcal{A}_{G}^{0}\circ\mathcal{E}_{T}\right)
(\rho),
\end{equation}
i.e., that any phase-insensitive Gaussian channel can be realized as a
concatenation of a pure-loss channel $\mathcal{E}_{T}$ of transmissivity $T$
followed by a quantum-limited amplifier channel $\mathcal{A}_{G}$ with gain
$G$, with $\tau=TG$ and $\nu=G\left(  1-T\right)  +G-1$. Thus, a photon number
state $\left\vert k\right\rangle \left\langle k\right\vert $\ input to the
phase-insensitive noise channel leads to an output of the following form:%
\begin{equation}
\mathcal{P}\left(  \left\vert k\right\rangle \left\langle k\right\vert
\right)  =\sum_{m=0}^{k}p\left(  m\right)  \mathcal{A}_{G}^{0}\left(
\left\vert m\right\rangle \left\langle m\right\vert \right)  ,
\label{amplifier-to-phaseinsensitive}%
\end{equation}
where
\begin{equation}
p\left(  m\right)  =\binom{k}{m}T^{m}\left(  1-T\right)  ^{k-m}.
\end{equation}
The quantum-limited amplifier channel has the following action on a photon
number state $\left\vert m\right\rangle $~\cite{PBLSC12}:%
\begin{equation}
\mathcal{A}_{G}^{0}\left(  \left\vert m\right\rangle \left\langle m\right\vert
\right)  =\sum_{l=0}^{\infty}q\left(  l|m\right)  \left\vert l\right\rangle
\left\langle l\right\vert ,
\end{equation}
where the conditional probabilities $q\left(  l|m\right)  $ are given by:%
\begin{equation}
q\left(  l|m\right)  =\left\{
\begin{array}
[c]{cc}%
0 & l<m\\
\left(  1-\mu^{2}\right)  ^{m+1}\mu^{2\left(  l-m\right)  }\binom{l}{l-m} &
l\geq m
\end{array}
\right.  ,
\end{equation}
where $\mu=\tanh r\in\left[  0,1\right]  $, with $r$ chosen such that
$G=\cosh^{2}\left(  r\right)  $.

The conditional distribution $q\left(  l|m\right)  $ has the two important
properties of having finite second moment and exponential decay with
increasing photon number. The property of exponential decay with increasing
$l$ follows from%
\begin{align}
\left(  1-\mu^{2}\right)  ^{m+1}\mu^{2\left(  l-m\right)  }\binom{l}{l-m}  &
=\left(  1-\mu^{2}\right)  ^{m+1}\mu^{-2m}2^{-2\log_{2}\left(  \frac{1}{\mu
}\right)  l}\binom{l}{l-m}\\
&  \leq\left(  1-\mu^{2}\right)  ^{m+1}\mu^{-2m}2^{-2\log_{2}\left(  \frac{1}{\mu
}\right)  l}2^{lh_{2}\left(  \frac{l-m}{l}\right)  }\\
&  =\left(  1-\mu^{2}\right)  ^{m+1}\mu^{-2m}2^{-l\left[  2\log_{2}\left(
\frac{1}{\mu}\right)  -h_{2}\left(  \frac{l-m}{l}\right)  \right]  }%
\end{align}
The inequality applies the bound $\binom{n}{k}\leq2^{nh_{2}\left(  k/n\right)
}$ (see (11.40)\ of \cite{CT06}), where $h_{2}\left(  x\right)  $ is the
binary entropy with the property that $\lim_{x\rightarrow1}h_{2}\left(
x\right)  =0$. Thus, for large enough $l$, it will be the case that
$2\log\left(  \frac{1}{\mu}\right)  -h_{2}\left(  \frac{l-m}{l}\right)  >0$,
so that the conditional distribution $q\left(  l|m\right)  $\ has exponential
decay with increasing $l$. We can also then conclude that this distribution
has a finite second moment. It follows from
(\ref{amplifier-to-phaseinsensitive}) that%
\begin{equation}
\mathcal{P}\left(  \left\vert k\right\rangle \left\langle k\right\vert
\right)  =\sum_{l=0}^{\infty}\left[  \sum_{m=0}^{k}p\left(  m\right)  q\left(
l|m\right)  \right]  \left\vert l\right\rangle \left\langle l\right\vert .
\label{eq:cond-dist-photon-num}%
\end{equation}

The eigenvalues above (i.e., $\sum_{m=0}^{k}p\left(  m\right)  q\left(
l|m\right)  $ ) represent a distribution over photon number states at the
output of the phase-insensitive channel $\mathcal{P}$, which we can write as a
conditional probability distribution $p\left(  l|k\right)  $ over $l$ given
the input with definite photon number~$k$. This probability distribution has
its mean equal to $\tau k+\left(  \tau+\nu-1\right)  /2$, since the mean
energy of the input state is $k$. Furthermore, this distribution inherits the
properties of having a finite second moment and an exponential decay to zero
as $l\rightarrow\infty$.

For example, we can consider the thermal noise channel $\mathcal{E}%
_{\eta,N_{B}}$ with the structural decomposition given by
(\ref{thermal-decomposition})%
\begin{equation}
\mathcal{E}_{\eta,N_{B}}(\rho)=(\mathcal{A}_{(1-\eta)N_{B}+1}^{0}%
\circ\mathcal{E}_{\eta/\left(  (1-\eta)N_{B}+1\right)  })(\rho).
\end{equation}
The mean of the corresponding distribution for this channel when a state of
definite photon number $k$ is input, following the above arguments, is equal
to $\eta k+\left(  1-\eta\right)  N_{B}$.

The argument from here is now exactly the same as the proof of Lemma~1 of
\cite{BW13} (starting from (40) of \cite{BW13}). We include it here for
completeness. We now suppose that the input state satisfies the
photon-number occupation constraint in (\ref{Constraint}), and apply the Gentle
Measurement Lemma~\cite{ON07,W99} to obtain the following inequality
\begin{equation}
\text{Tr}\left\{  \Pi_{\left\lceil nN_{S}^{\prime}+\delta_{2}\right\rceil
}\mathcal{P}^{\otimes n}\left(  \rho^{\left(  n\right)  }\right)  \right\}
\geq\text{Tr}\left\{  \Pi_{\left\lceil nN_{S}^{\prime}+\delta_{2}\right\rceil
}\mathcal{P}^{\otimes n}\left(  \Pi_{\left\lceil nN_{S}\right\rceil }%
\rho^{\left(  n\right)  }\Pi_{\left\lceil nN_{S}\right\rceil }\right)
\right\}  -2\sqrt{\delta_{1}(n)}, \label{eq:photon-counting-output}%
\end{equation}
where $N_{S}^{\prime}=\tau N_{S}+\left(  \tau+\nu-1\right)  /2$. Since there
is photodetection at the output (i.e., the projector $\Pi_{\left\lceil n\eta
N_{S}^{\prime}+\delta_{2}\right\rceil }$ is diagonal in the number basis), it
suffices for us to consider the input $\Pi_{\left\lceil nN_{S}\right\rceil
}\rho^{\left(  n\right)  }\Pi_{\left\lceil nN_{S}\right\rceil }$ to be
diagonal in the photon-number basis, and we write this as%
\begin{equation}
\rho^{\left(  n\right)  }=\sum_{a^{n}:\sum_{i}a_{i}\leq\lceil nN_{S}\rceil
}p\left(  a^{n}\right)  \left\vert a^{n}\right\rangle \left\langle
a^{n}\right\vert ,
\end{equation}
where $\left\vert a^{n}\right\rangle $ represents strings of photon number
states. We then find that (\ref{eq:photon-counting-output}) is equal to%
\begin{multline}
  \sum_{a^{n}:\sum_{i}a_{i}\leq\left\lceil nN_{S}\right\rceil }p\left(
a^{n}\right)  \text{Tr}\left\{  \left(  \Pi_{\left\lceil nN_{S}^{\prime
}+\delta_{2}\right\rceil }\right)  \mathcal{P}^{\otimes n}\left(  \left\vert
a^{n}\right\rangle \left\langle a^{n}\right\vert \right)  \right\}
-2\sqrt{\delta_{1}(n)}\\
  =\sum_{a^{n}:\sum_{i}a_{i}\leq\left\lceil nN_{S}\right\rceil }p\left(
a^{n}\right)  \sum_{l^{n}:\sum_{i}l_{i}\leq\left\lceil nN_{S}^{\prime}%
+\delta_{2}\right\rceil }p\left(  l^{n}|a^{n}\right)  -2\sqrt{\delta_{1}(n)},
\label{analysisterm}%
\end{multline}
where the distribution $p\left(  l^{n}|a^{n}\right)  \equiv\prod
\limits_{i=1}^{n}p\left(  l_{i}|a_{i}\right)  $ with $p\left(  l_{i}%
|a_{i}\right)  $ coming from \eqref{eq:cond-dist-photon-num}.

In order to obtain a lower bound on the expression in (\ref{analysisterm}), we
analyze the term
\begin{equation}
\sum_{l^{n}:\sum_{i}l_{i}\leq\left\lceil nN_{S}^{\prime}+\delta_{2}%
\right\rceil }p\left(  l^{n}|a^{n}\right)  \label{conditional}%
\end{equation}
on its own under the assumption that $\sum_{i}a_{i}\leq\left\lceil
nN_{S}\right\rceil $. Let $L_{i}|a_{i}$ denote a conditional random variable
with distribution $p\left(  l_{i}|a_{i}\right)  $, and let $\overline{L^{n}%
}|a^{n}$ denote the sum random variable:%
\begin{equation}
\overline{L^{n}}|a^{n}\equiv\sum_{i}L_{i}|a_{i},
\end{equation}
so that%
\begin{align}
\sum_{l^{n}:\sum_{i}l_{i}\leq\left\lceil n N_{S}^{\prime}+\delta_{2}%
\right\rceil }p\left(  l^{n}|a^{n}\right)   &  =\Pr\left\{  \overline{L^{n}%
}|a^{n}\leq n(N_{S}^{\prime}+\delta_{2})\right\} \\
&  =\Pr\left\{  \overline{L^{n}}|a^{n}\leq n\left(  \tau N_{S}+\left(
\tau+\nu-1\right)  /2 +\delta_{2} \right )\right\} \\
&  \geq\Pr\left\{  \overline{L^{n}}|a^{n}\leq n\left(  \tau\frac{1}{n}\sum
_{i}a_{i}+\left(  \tau+\nu-1\right)  /2+\delta_{2}\right)  \right\}  ,
\label{eq:probability-to-bound}%
\end{align}
where $\left(  \tau+\nu-1\right)  /2$ represents the mean number of noise
photons injected by the channel, and the inequality follows from the
constraint $\sum_{i}a_{i}\leq\left\lceil nN_{S}\right\rceil $. Since%
\begin{equation}
\mathbb{E}\left\{  L_{i}|a_{i}\right\}  =\tau a_{i}+\left(  \tau+\nu-1\right)
/2,
\end{equation}
it follows that%
\begin{equation}
\mathbb{E}\left\{  \overline{L^{n}}|a^{n}\right\}  =n\left(  \tau\frac{1}%
{n}\sum_{i}a_{i}+\left(  \tau+\nu-1\right)  /2\right)  ,
\end{equation}
and so the expression in (\ref{eq:probability-to-bound}) is the probability
that a sum of independent random variables deviates from its mean by no more
than $\delta_{2}$. To obtain a bound on the probability in
(\ref{eq:probability-to-bound}) from below, we now follow the approach in
\cite{BW13} employing the truncation method (see Section~2.1 of \cite{T12} for
more details), in which each random variable $L_{i}|a_{i}$ is split into two
parts:%
\begin{align}
\left(  L_{i}|a_{i}\right)  _{>T_{0}}  &  \equiv\left(  L_{i}|a_{i}\right)
\mathcal{I}\left(  \left(  L_{i}|a_{i}\right)  >T_{0}\right)  ,\\
\left(  L_{i}|a_{i}\right)  _{\leq T_{0}}  &  \equiv\left(  L_{i}%
|a_{i}\right)  \mathcal{I}\left(  \left(  L_{i}|a_{i}\right)  \leq
T_{0}\right)  ,
\end{align}
where $\mathcal{I}\left(  \cdot\right)  $ is the indicator function and
$T_{0}$ is a truncation parameter taken to be very large (much larger than
$\max_{i}a_{i}$, for example). We can then split the sum random variable into
two parts as well:%
\begin{align}
\overline{L^{n}}|a^{n}  &  =\left(  \overline{L^{n}}|a^{n}\right)  _{>T_{0}%
}+\left(  \overline{L^{n}}|a^{n}\right)  _{\leq T_{0}}\\
&  \equiv\sum_{i}\left(  L_{i}|a_{i}\right)  _{>T_{0}}+\sum_{i}\left(
L_{i}|a_{i}\right)  _{\leq T_{0}}.
\end{align}
We can use the union bound to argue that%
\begin{multline}
\Pr\left\{  \overline{L^{n}}|a^{n}\geq\mathbb{E}\left\{  \overline{L^{n}%
}|a^{n}\right\}  +n\delta_{2}\right\}  \leq\Pr\left\{  \left(  \overline
{L^{n}}|a^{n}\right)  _{>T_{0}}\geq\mathbb{E}\left\{  \left(  \overline{L^{n}%
}|a^{n}\right)  _{>T_{0}}\right\}  +n\delta_{2}/2\right\} \\
+\Pr\left\{  \left(  \overline{L^{n}}|a^{n}\right)  _{\leq T_{0}}%
\geq\mathbb{E}\left\{  \left(  \overline{L^{n}}|a^{n}\right)  _{\leq T_{0}%
}\right\}  +n\delta_{2}/2\right\}  .
\end{multline}
The idea from here is that for a random variable $L_{i}|a_{i}$ with sufficient
decay for large values, we can bound the first probability for $\left(
\overline{L^{n}}|a^{n}\right)  _{>T_{0}}$ from above by $\varepsilon
/\delta_{2}$\ for $\varepsilon$ an arbitrarily small positive constant (made
small by taking $T_{0}$ larger)\ by employing the Markov inequality. We then
bound the second probability for $\left(  \overline{L^{n}}|a^{n}\right)
_{\leq T_{0}}$ using a Chernoff bound, since these random variables are
bounded. This latter bound has an exponential decay with increasing $n$ due to
the ability to use a Chernoff bound. So, the argument is just to make
$\varepsilon$ arbitrarily small by increasing the truncation parameter $T_{0}%
$, and for $n$ large enough, exponential convergence to zero kicks in. We
point the reader to Section~2.1 of \cite{T12} for more details. By using
either approach, we arrive at the following bound:%
\begin{equation}
\sum_{l^{n}:\sum_{i}l_{i}\leq\left\lceil nN_{S}^{\prime}+\delta_{2}%
\right\rceil }p\left(  l^{n}|a^{n}\right)  \geq1-\delta_{3}(n),
\end{equation}
where $\delta_{3}(n)$ is a function decreasing to zero as $n\rightarrow\infty
$. Finally, we put this together with (\ref{analysisterm}) to obtain%
\begin{align}
&  \text{Tr}\left\{  \Pi_{\left\lceil nN_{S}^{\prime}+\delta_{2}\right\rceil
}\mathcal{P}^{\otimes n}\left(  \rho^{\left(  n\right)  }\right)  \right\} \\
&  \geq\sum_{a^{n}:\sum_{i}a_{i}\leq\left\lceil nN_{S}\right\rceil }p\left(
a^{n}\right)  \sum_{l^{n}:\sum_{i}l_{i}\leq\left\lceil nN_{S}^{\prime}%
+\delta_{2}\right\rceil }p\left(  l^{n}|a^{n}\right)  -2\sqrt{\delta_{1}(n)}\\
&  \geq\left(  1-\delta_{1}(n)\right)  \left(  1-\delta_{3}(n)\right)
-2\sqrt{\delta_{1}(n)}\\
&  \geq1-\delta_{1}(n)-\delta_{3}(n)-2\sqrt{\delta_{1}(n)},
\end{align}
thereby completing the proof.
\end{IEEEproof}

\bigskip

Let $\Lambda_{m}$ denote a decoding POVM acting on the output space of $n$
instances of the phase-insensitive channel. In what follows, we prove the
strong converse theorem for the classical capacity of all phase-insensitive
Gaussian channels.
\bigskip
\begin{theorem}
\label{Theorem}Let $\mathcal{P}$ be a phase-insensitive Gaussian channel with
parameters $\tau$ and $\nu$ as defined in
\eqref{eq:phase-insensitive-Gaussian}. The average success probability
$p_{\text{succ}}$ of any code for this channel satisfying \eqref{Constraint}
is bounded as%
\begin{multline}
p_{\operatorname{succ}}=\frac{1}{M}\sum_{m}\operatorname{Tr}\{\Lambda
_{m}\mathcal{P}^{\otimes n}(\rho_{m})\}\leq
2^{-nR}2^{n\left[  g\left(  N_{S}^{\prime}\right)  -H_{\alpha}(\mathcal{P}%
(|0\rangle\langle0|))+\delta_{2}+\tfrac{1}{n(\alpha-1)}\log_{2}(1/\varepsilon
)\right]  }+\varepsilon+\delta_{6}(n),
\end{multline}
where $\alpha>1$, $\varepsilon\in(0,1)$, $N_{S}^{\prime}=\tau N_{S}+\left(
\tau+\nu-1\right)  /2$, $\mathcal{P}^{\otimes n}$ denotes $n$ instances of
$\mathcal{P}$, $\delta_{1}(n)$ is defined in \eqref{Constraint}, $\delta_{2}$
is an arbitrarily small positive constant, $\delta_{3}(n)$ is a function
decreasing with $n$ (both defined in Lemma~\ref{lem:photon-num-lemma}),
and $\delta_{6}(n)=2\sqrt{\delta_{1}(n)+2\sqrt{\delta_{1}(n)}+\delta_{3}(n)}$. 
\end{theorem}

\begin{IEEEproof}
This proof is very similar to the proof of Theorem~2 of \cite{BW13}, with the
exception that we can now invoke the main result of \cite{MGH13}\ (that the
minimum output entropy for R\'enyi entropies of arbitrary order is attained by
the vacuum state input). Consider the success probability of any code
satisfying the photon-number occupation constraint (\ref{Constraint}):%
\begin{align}
\frac{1}{M}\sum_{m}\text{Tr}\{\Lambda_{m}\mathcal{P}^{\otimes n}(\rho_{m})\}
&  \leq\frac{1}{M}\sum_{m}\text{Tr}\{\Lambda_{m}\Pi_{\lceil nN_{S}^{\prime
}\rceil}\mathcal{P}^{\otimes n}(\rho_{m})\Pi_{\lceil nN_{S}^{\prime}\rceil
}\} \nonumber\\
&  \ \ \ \ \ \ +\frac{1}{M}\sum_{m}\left\Vert \Pi_{\lceil nN_{S}^{\prime
}\rceil}\mathcal{P}^{\otimes n}(\rho_{m})\Pi_{\lceil nN_{S}^{\prime}\rceil
}-\mathcal{P}^{\otimes n}(\rho_{m})\right\Vert _{1}\\
&  \leq\frac{1}{M}\sum_{m}\text{Tr}\{\Lambda_{m}\Pi_{\lceil nN_{S}^{\prime
}\rceil}\mathcal{P}^{\otimes n}(\rho_{m})\Pi_{\lceil nN_{S}^{\prime}\rceil
}\}\nonumber\\
&  \ \ \ \ \ \ +2\sqrt{\delta_{1}(n)+2\sqrt{\delta_{1}(n)}+\delta_{3}(n)}.
\end{align}
The first inequality is a special case of the inequality
\begin{equation}
\operatorname{Tr}\{\Lambda\sigma\}\leq\text{Tr}\{\Lambda\rho\}+\left\Vert
\rho-\sigma\right\Vert _{1}, \label{ineq1}%
\end{equation}
which holds for $0\leq\Lambda\leq I$, $\rho,\sigma\geq0$, and
$\operatorname{Tr}\{\rho\},\operatorname{Tr}\{\sigma\}\leq1$. The second
inequality is obtained by invoking Lemma~\ref{lem:photon-num-lemma} and the
Gentle Measurement Lemma~\cite{ON07,W99} for ensembles.

Note that in the above, the second term vanishes as $n\rightarrow\infty$;
hence it suffices to focus on the first term, which by cyclicity of trace
yields%
\begin{equation}
\frac{1}{M}\sum_{m}\text{Tr}\{\Lambda_{m}\Pi_{\lceil nN_{S}^{\prime}\rceil
}\mathcal{P}^{\otimes n}(\rho_{m})\Pi_{\lceil nN_{S}^{\prime}\rceil}%
\}=\frac{1}{M}\sum_{m}\text{Tr}\{\Pi_{\lceil nN_{S}^{\prime}\rceil}\Lambda
_{m}\Pi_{\lceil nN_{S}^{\prime}\rceil}\mathcal{P}^{\otimes n}(\rho_{m})\}.
\label{eq:succ-prob-projected}%
\end{equation}

At this point, we consider the set of all states $\widetilde{\sigma}_{m}$ that
are $\varepsilon$-close in trace distance to each output of the
phase-insensitive channel $\mathcal{P}^{\otimes n}\left(  \rho_{m}\right)  $
(let us denote this set by $\mathcal{B}^{\varepsilon}\left(  \mathcal{P}%
^{\otimes n}\left(  \rho_{m}\right)  \right)  $. This consideration will allow
us to relate the success probability to the smooth min-entropy. We find the
following upper bound on (\ref{eq:succ-prob-projected}):%
\begin{align}
\frac{1}{M}\sum_{m}\text{Tr}\{\Pi_{\lceil nN_{S}^{\prime}\rceil}\Lambda_{m}%
\Pi_{\lceil nN_{S}^{\prime}\rceil}\mathcal{P}^{\otimes n}(\rho_{m})\}  &
\leq\frac{1}{M}\sum_{m}\text{Tr}\{\Pi_{\lceil nN_{S}^{\prime}\rceil}%
\Lambda_{m}\Pi_{\lceil nN_{S}^{\prime}\rceil}\widetilde{\sigma}_{m}%
\}+\varepsilon\\
&  \leq\frac{1}{M}\sum_{m}\text{Tr}\{\Pi_{\lceil nN_{S}^{\prime}\rceil}%
\Lambda_{m}\Pi_{\lceil nN_{S}^{\prime}\rceil}\}\left\Vert \widetilde{\sigma
}_{m}\right\Vert _{\infty}+\varepsilon.
\end{align}
We can now optimize over all of the states $\widetilde{\sigma}_{m}$ that are
$\varepsilon$-close to $\mathcal{P}^{\otimes n}\left(  \rho_{m}\right)  $,
leading to the tightest upper bound on the success probability%
\begin{multline}
  \frac{1}{M}\sum_{m}\text{Tr}\{\Pi_{\lceil nN_{S}^{\prime}\rceil}\Lambda
_{m}\Pi_{\lceil nN_{S}^{\prime}\rceil}\mathcal{P}^{\otimes n}(\rho_{m})\}\\
  \leq\frac{1}{M}\sum_{m}\text{Tr}\{\Pi_{\lceil nN_{S}^{\prime}\rceil}%
\Lambda_{m}\Pi_{\lceil nN_{S}^{\prime}\rceil}\}\inf_{\widetilde{\sigma}_{m}%
\in\mathcal{B}^{\varepsilon}\left(  \mathcal{P}^{\otimes n}\left(  \rho
_{m}\right)  \right)  }\left\Vert \widetilde{\sigma}_{m}\right\Vert _{\infty
}+\varepsilon. \label{ineq2}%
\end{multline}
Since the quantity $\inf_{\widetilde{\sigma}_{m}\in\mathcal{B}^{\varepsilon
}\left(  \mathcal{P}^{\otimes n}\left(  \rho_{m}\right)  \right)  }\left\Vert
\widetilde{\sigma}_{m}\right\Vert _{\infty}$ is related to the smooth
min-entropy via%
\begin{equation}
\inf_{\widetilde{\sigma}_{m}\in\mathcal{B}^{\varepsilon}\left(  \mathcal{P}%
^{\otimes n}\left(  \rho_{m}\right)  \right)  }\left\Vert \widetilde{\sigma
}_{m}\right\Vert _{\infty}=2^{-H_{\min}^{\varepsilon}\left(  \mathcal{P}%
^{\otimes n}(\rho_{m})\right)  },
\end{equation}
the upper bound in (\ref{ineq2}) gives
\begin{align}
&  \frac{1}{M}\sum_{m}\text{Tr}\{\Pi_{\lceil nN_{S}^{\prime}\rceil}\Lambda
_{m}\Pi_{\lceil nN_{S}^{\prime}\rceil}\}2^{-H_{\min}^{\varepsilon}\left(
\mathcal{P}^{\otimes n}(\rho_{m})\right)  }+\varepsilon\nonumber\\
&  \leq\frac{1}{M}\sum_{m}\text{Tr}\{\Pi_{\lceil nN_{S}^{\prime}\rceil}%
\Lambda_{m}\Pi_{\lceil nN_{S}^{\prime}\rceil}\}\sup_{\rho}2^{-H_{\min
}^{\varepsilon}\left(  \mathcal{P}^{\otimes n}(\rho)\right)  }+\varepsilon
\nonumber\\
&  =\frac{1}{M}2^{-\inf_{\rho}H_{\min}^{\varepsilon}\left(  \mathcal{P}%
^{\otimes n}(\rho)\right)  }\text{Tr}\{\Pi_{\lceil nN_{S}^{\prime}\rceil
}\}+\varepsilon\nonumber\\
&  \leq2^{-nR}2^{-\inf_{\rho}H_{\min}^{\varepsilon}\left(  \mathcal{P}%
^{\otimes n}(\rho)\right)  }2^{n\left[  g\left(  N_{S}^{\prime}\right)
+\delta\right]  }+\varepsilon.
\end{align}
The first inequality follows by taking a supremum over all input states. The
first equality follows because $\sum_{m}\Lambda_{m}=I$ for the set of decoding
POVM measurements $\{\Lambda_{m}\}$, and the second inequality is a result of
the upper bound on the rank of the photon number cutoff projector in
(\ref{Rank}). We have also used the fact that the rate of the channel is
expressed as $R=(\log_{2}M)/n$, where $M$ is the number of messages.

Observe that the success probability is now related to the smooth min-entropy,
and we can exploit the following relation between smooth min-entropy and
the R\'{e}nyi entropies for $\alpha>1$ \cite{RWISIT1}:%
\begin{equation}
H_{\min}^{\varepsilon}\left(  \omega\right)  \geq H_{\alpha}\left(
\omega\right)  -\frac{1}{\alpha-1}\log_{2}\left(  \frac{1}{\varepsilon}\right)  .
\end{equation}
Using the above inequality and the fact that the \textquotedblleft
strong\textquotedblright\ Gaussian optimizer conjecture has been proven for
the R\'{e}nyi entropies of all orders~\cite{MGH13} (recall
\eqref{eq:main-result-MGH}), we get that%
\begin{equation}
\inf_{\rho}H_{\min}^{\varepsilon}\left(  \mathcal{P}^{\otimes n}(\rho)\right)
\geq n\left[  H_{\alpha}\left(  \mathcal{P}(\left\vert 0\right\rangle
\left\langle 0\right\vert )\right)  -\frac{1}{n\left(  \alpha-1\right)  }%
\log_{2}\left(  \frac{1}{\varepsilon}\right)  \right]  . \label{ineq3}%
\end{equation}
The first term on the right hand side is a result of the fact that the vacuum
state minimizes the R\'{e}nyi entropy of all orders at the output of a
phase-insensitive Gaussian channel.
\end{IEEEproof}

\bigskip

By tuning the parameters $\alpha$ and $\varepsilon$ appropriately, we recover
the strong converse theorem:

\bigskip

\begin{corollary}
[Strong converse]\label{cor:main-result}Let $\mathcal{P}$ be a
phase-insensitive Gaussian channel with parameters $\tau$ and $\nu$ as defined
in \eqref{eq:phase-insensitive-Gaussian}. The average success probability
$p_{\text{succ}}$ of any code for this channel satisfying \eqref{Constraint}
is bounded as
\begin{multline}
p_{\operatorname{succ}}=\frac{1}{M}\sum_{m}\operatorname{Tr}\{\Lambda
_{m}\mathcal{P}^{\otimes n}(\rho_{m})\}\leq 2^{-nR}2^{n\left[  g\left(  N_{S}^{\prime}\right)
 -g\left(  N_{B}^{\prime
}\right)  +\delta_{2}+\delta_{5}/\delta_{4}+\delta_{4}K\left(  N_{B}^{\prime
}\right)  \right]  }+2^{-n\delta_{5}}+\delta_{6}(n),
\end{multline}
where $N_{S}^{\prime}=\tau N_{S}+\left(  \tau+\nu-1\right)  /2$,
$N_{B}^{\prime}\equiv\left(  \tau+\nu-1\right)  /2$, $\mathcal{P}^{\otimes n}$
denotes $n$ instances of $\mathcal{P}$,
$\delta_{1}(n)$ is defined in \eqref{Constraint}, $\delta_{2}$ is an arbitrarily small positive constant, 
$\delta_{3}(n)$ is a function decreasing with $n$ (both defined in
Lemma~\ref{lem:photon-num-lemma}), $\delta_{4}$ and $\delta_{5}$ are
arbitrarily small positive constants such that $\delta_{5}/\delta_{4}$ is
arbitrarily small, and $K\left(  N_{B}^{\prime}\right)  $ is a function of
$N_{B}^{\prime}$ only. Also, $\delta_{6}(n)=2\sqrt{\delta_{1}(n)+2\sqrt{\delta_{1}(n)}+\delta_{3}(n)}$.
Thus, for any rate $R>g\left(  N_{S}^{\prime}\right)
-g\left(  N_{B}^{\prime}\right)  $, it is possible to choose the parameters
such that the success probability of any family of codes satisfying
\eqref{Constraint} decreases to zero in the limit of large~$n$.
\end{corollary}

\begin{IEEEproof}
In Theorem~\ref{Theorem}, we can pick $\alpha=1+\delta_{4}$ and $\varepsilon
=2^{-n\delta_{5}}$, with $\delta_{5}>0$ much smaller than $\delta_{4}>0$ such
that $\delta_{5}/\delta_{4}$ is arbitrarily small, and the terms on the right
hand side in (\ref{ineq3}) simplify to%
\begin{equation}
n\left[  H_{1+\delta_{4}}\left(  \mathcal{P}(\left\vert 0\right\rangle
\left\langle 0\right\vert )\right)  -\frac{\delta_{5}}{\delta_{4}}\right]  .
\end{equation}
The output state $\mathcal{P}(\left\vert 0\right\rangle \left\langle
0\right\vert )$ for the phase-insensitive channel with the vacuum state as the
input is a thermal state with mean photon number $N_{B}^{\prime}\equiv\left(
\tau+\nu-1\right)  /2$. The quantum R\'{e}nyi entropy of order $\alpha>1$ of a
thermal state with mean photon number $N_{B}^{\prime}$ is given
by~\cite{GGLMS04}%
\begin{equation}
\frac{\log_{2}\left[  \left(  N_{B}^{\prime}+1\right)  ^{\alpha}-N_{B}%
^{\prime\alpha}\right]  }{\alpha-1}.\label{eq:renyi-thermal}%
\end{equation}
Lemma~6.3 of~\cite{Tomamichelthesis} gives us the following inequality for a
general state (for $\alpha$ close enough to one):%
\begin{equation}
H_{\alpha}\left(  \rho\right)  \geq H\left(  \rho\right)  -4\left(
\alpha-1\right)  \left(  \log_{2} v\right)  ^{2},
\end{equation}
where%
\begin{equation}
v\equiv2^{-\frac{1}{2}H_{3/2}\left(  \rho\right)  }+2^{\frac{1}{2}%
H_{1/2}\left(  \rho\right)  }+1 .
\end{equation}
For a thermal state, we find using\ (\ref{eq:renyi-thermal}) that%
\begin{align}
H_{3/2}\left(  \rho\right)    & =2\log_{2}\left[  \left(  N_{B}^{\prime}+1\right)
^{3/2}-N_{B}^{\prime3/2}\right]  ,\\
H_{1/2}\left(  \rho\right)    & =-2\log_{2}\left[  \left(  N_{B}^{\prime
}+1\right)  ^{1/2}-N_{B}^{\prime1/2}\right]  ,
\end{align}
so that%
\begin{equation}
v\left(  N_{B}^{\prime}\right)  =\left[  \left(  N_{B}^{\prime}+1\right)
^{3/2}-N_{B}^{\prime3/2}\right]  ^{2}+\left[  \left(  N_{B}^{\prime}+1\right)
^{1/2}-N_{B}^{\prime1/2}\right]  ^{-2}+1.
\end{equation}
We then find that%
\begin{align}
H_{1+\delta_{4}}\left(  \mathcal{P}(\left\vert 0\right\rangle \left\langle
0\right\vert )\right)    & \geq H\left(  \mathcal{P}(\left\vert 0\right\rangle
\left\langle 0\right\vert )\right)  -\delta_{4}K\left(  N_{B}^{\prime}\right)
\\
& =g\left(  N_{B}^{\prime}\right)  -\delta_{4}K\left(  N_{B}^{\prime}\right)
,
\end{align}
where%
\begin{equation}
K\left(  N_{B}^{\prime}\right)  \equiv4\left[  \log_{2} v\left(  N_{B}^{\prime
}\right)  \right]  ^{2}.
\end{equation}
We now recover the bound in the statement of the corollary.
\end{IEEEproof}

Finally, we recall the capacities of the phase-insensitive channels in
(\ref{thermalcapacity}), (\ref{additivecapacity}), and
(\ref{amplifiercapacity}). Comparing them with the statement of
Corollary~\ref{cor:main-result}, we can conclude that these expressions indeed
represent strong converse rates for these respective channels, since the
success probability when communicating above these rates decreases to zero in
the limit $n\rightarrow\infty$.

\section{Conclusion}

\label{sec:conclusion}Phase-insensitive Gaussian channels represent physical
noise models which are relevant for optical communication, including
attenuation, thermalization, or amplification of optical signals.
In this paper, we combine the proofs in \cite{BW13} with the recent results of
\cite{GHG13,MGH13,GPCH13} to prove that a strong converse theorem holds for the
classical capacity of all phase-insensitive Gaussian quantum channels. We
showed that the success probability of correctly decoding classical
information asymptotically converges to zero in the limit of many channel
uses, if the communication rate exceeds the capacity. Our result thus
establishes the capacity of these channels as a very sharp dividing line
between possible and impossible communication rates through these channels.
This result might find an immediate application in proving security of noisy
quantum storage models of cryptography~\cite{KS4} for continuous-variable systems.
The results of this paper can also be easily extended to the more general case of 
multimode bosonic Gaussian channels~\cite{GHG13}.

As an open question, one might attempt to establish a strong converse for the
classical capacity of all \textit{phase-sensitive} Gaussian channels. Another
area of research where our result might be extended is in the setting of
network information theory---for example, one might consider establishing a
strong converse for the classical capacity of the multiple-access bosonic
channels, in which two or more senders communicate to a common receiver over a
shared communication channel~\cite{MAC}.

\bibliographystyle{IEEEtran}
\bibliography{Ref}

\begin{IEEEbiographynophoto}{Bhaskar Roy Bardhan} was born in Kolkata, West Bengal, India.
He received the B.~Sc.~degree in physics (honors) from St. Xavier's College, University of Calcutta, India, in 
2006, the M.~Sc.~degree in physics from Indian Institute of Technology Guwahati, India, 
in 2008, and the Ph.D. degree in physics from Louisiana State University, Baton Rouge, 

Currently, he is a Postdoctoral Associate at the Research Laboratory of Electronics, 
Massachusetts Institute of Technology, Cambridge, Massachusetts. His research interests 
include optical and quantum communication, quantum information theory, and quantum 

Dr. Roy Bardhan is a member of the American Physical Society, the Optical Society of 
America, and IEEE, and has been a reviewer for the journals Physical Review Letters, 
Physical Review A, and Optics Letters.
\end{IEEEbiographynophoto}

\begin{IEEEbiographynophoto}{Raul Garcia-Patron} was born in Madrid, Spain. He received the Ph.D. degree in engineering from the Universit\'e Libre de Bruxelles, Brussels, Belgium, in 2007. He has previously hold postdoctoral 
positions at Massachusetts Institute of Technology and Max-Planck Institute for Quantum Optics. He 
currently benefits from a BELSPO return fellowship, funded by the Belgian Federal Government, in 
the Centre for Quantum Information and Communication at Universit\'e Libre de Bruxelles. His 
current research interests are in quantum Shannon theory and quantum information processing 
with photonic systems.
\end{IEEEbiographynophoto}

\begin{IEEEbiographynophoto}{Mark M. Wilde} (M'99-SM'13) was born in Metairie, Louisiana, USA. He received the Ph.D. degree in electrical engineering from the University of Southern California, Los Angeles, California, in 2008. He is an Assistant Professor in the Department of Physics and Astronomy and the Center for Computation and Technology at Louisiana State University. His current research interests are in quantum Shannon theory, quantum optical communication, quantum computational complexity theory, and quantum error correction.
\end{IEEEbiographynophoto}

\begin{IEEEbiographynophoto}{Andreas Winter} received a Diploma degree in Mathematics from the Freie
Universit\"at Berlin, Berlin, Germany, in 1997, and a Ph.D. degree from the
Fakult\"at f\"ur Mathematik, Universit\"at Bielefeld, Bielefeld, Germany, in 1999.
He was Research Associate at the University of Bielefeld until 2001, and then
with the Department of Computer Science at the University of Bristol, Bristol,
UK. In 2003, still with the University of Bristol, he was appointed Lecturer
in Mathematics, and in 2006 Professor of Physics of Information. Since 2012
he has been an ICREA Research Professor with the Universitat Aut\`onoma de
Barcelona, Barcelona, Spain.
\end{IEEEbiographynophoto}

\end{document}